\documentclass{elsart}
\usepackage{amsfonts}
\usepackage{amssymb}
\usepackage{amsmath}
\usepackage[dvips]{graphicx}

\begin{document}

\begin{frontmatter}

\title{Breaking parameter modulated chaotic secure communication system}

\author{G. \'{A}lvarez\corauthref{corr}},
\author{F. Montoya},
\author{M. Romera},
\author{G. Pastor}

\corauth[corr]{Corresponding author: Email: gonzalo@iec.csic.es}

\address{Instituto de F\'{\i}sica Aplicada, Consejo Superior de
Investigaciones Cient\'{\i}ficas, Serrano 144---28006 Madrid,
Spain}

\begin{abstract}
This paper describes the security weakness of a recently proposed
secure communication method based on parameter modulation of a
chaotic system and adaptive observer-based synchronization scheme.
We show that the security is compromised even without precise
knowledge of the chaotic system used.

\end{abstract}

\end{frontmatter}

\section{Introduction}

In recent years, a growing number of cryptosystems based on chaos
have been proposed ~\cite{Yang}, many of them fundamentally flawed
by a lack of robustness and security. In \cite{Feki}, the author
proposes a symmetric secure communication system based on
parameter modulation of a chaotic oscillator acting as a
transmitter. The receiver is a chaotic system synchronized by
means of an adaptive observer.

In this paper we show how to break the proposed cryptosystem when
Lorenz's attractor is used as the non-linear time-varying
system~(\cite[\S 3.2]{Feki}), which, in fact, was the only example
explained in detail. Lorenz system is described by the following
equations:
\begin{align}
   \dot{x}_{1}&=-\sigma_{1}x_{1}+\sigma_{2}x_{2}, \label{eq:alpha}\\
   \dot{x}_{2}&=rx_{1}-x_{2}-x_{1}x_{3},  \label{eq:beta}\\
   \dot{x}_{3}&=x_{1}x_{2}-bx_{3}. \label{eq:delta}
\end{align}

In the example the system is implemented with the following
parameter values, $(\sigma_1, \sigma_2,r,b)=(10, 10, 28, 8/3)$.
The signal used for synchronization of the receiver is $x_1$. The
encryption process is defined by modulating the parameter
$\sigma_1$ with the binary encoded plaintext, so that it is
$\sigma_1+2.5$ if the plaintext bit is "1" and $\sigma_1-2.5$ if
the plaintext bit is "0". The duration of the plaintext bits must
be much larger than the convergence time of the adaption law.
Actually, in the example the bit rate is 0.2 bits/second. The
uncertain system can be rewritten in a compact form as:

\begin{equation}\label{eq:matrix1}
\left[ {\begin{array}{*{20}c}
   {\dot x_1 }  \\
   {\dot x_2 }  \\
   {\dot x_3 }  \\
\end{array}} \right] = \left[ {\begin{array}{*{20}c}
   { - \sigma_1} & {\sigma_2} & 0  \\
   {r} & { - 1} & 0  \\
   0 & 0 & -b  \\
\end{array}} \right]\left[ {\begin{array}{*{20}c}
   {x{}_1}  \\
   {x{}_2}  \\
   {x{}_3}  \\
\end{array}} \right] + \left( {\begin{array}{*{20}c}
   0  \\
   { - x{}_1x{}_3}  \\
   {x{}_1x{}_2}  \\
\end{array}} \right) + \left[ {\begin{array}{*{20}c}
   1  \\
   0  \\
   0  \\
\end{array}} \right]( - y)\theta
\end{equation}

\begin{equation}\label{eq:y}
y=C\cdot x=x_1
\end{equation}
\begin{equation}\label{eq:C}
C=[1~0~0]
\end{equation}
\begin{equation}\label{eq:teta}
\theta=\Delta\sigma_1=\pm 2.5
\end{equation}

The decryption process consists of a chaotic system synchronized
by means of an adaptive observer. The observer-based response
system is designed as:

\begin{equation}\label{eq:matrix2}
\left[ {\begin{array}{*{20}c}
   {\dot {\hat {x}}_1 }  \\
   {\dot {\hat {x}}_2 }  \\
   {\dot {\hat {x}}_3 }  \\
\end{array}} \right] = \left[ {\begin{array}{*{20}c}
   { - \sigma_1} & {\sigma_2} & 0  \\
   {r} & { - 1} & 0  \\
   0 & 0 & -b\\
\end{array}} \right]\left[ {\begin{array}{*{20}c}
   {\hat x{}_1}  \\
   {\hat x{}_2}  \\
   {\hat x{}_3}  \\
\end{array}} \right] + \left( {\begin{array}{*{20}c}
   0  \\
   { - \hat x{}_1\hat x{}_3}  \\
   {\hat x{}_1\hat x{}_2}  \\
\end{array}} \right) + LC(x_1  - \hat x_1 )
\end{equation}
\begin{equation}\label{eq:L}
L=[0~38~0]^{T}
\end{equation}

The plaintext can be retrieved from the first derivative of the
receiver uncertainty defined as:
\begin{equation}\label{eq:teta}
\dot{\hat{\theta}}=-5y(x_1  - \hat x_1 )
\end{equation}

The initial conditions of the transmitter and receiver are:
$(x_1(0),x_2(0),x_3(0))=(10,15,20)$ and
$(\hat{x}_1(0),\hat{x}_2(0),\hat{x}_3(0),\hat{\theta}(0))=(0,0,0,0)$.

Although the author seemed to base the security of its
cryptosystem on the chaotic behavior of the output of the Lorenz
non-linear system, no analysis of security was included. It was
not considered whether there should be a key in the proposed
system, what it should consist of, what the available key space
would be, and how it would be managed. We discuss the weaknesses
of this secure communication system in Sec.~\ref{sec:powerattack}
and in Sec.~\ref{sec:GSattack}.

\section{Power analysis attack}
\label{sec:powerattack} 
\begin{center}
\begin{figure}[h]
  \includegraphics{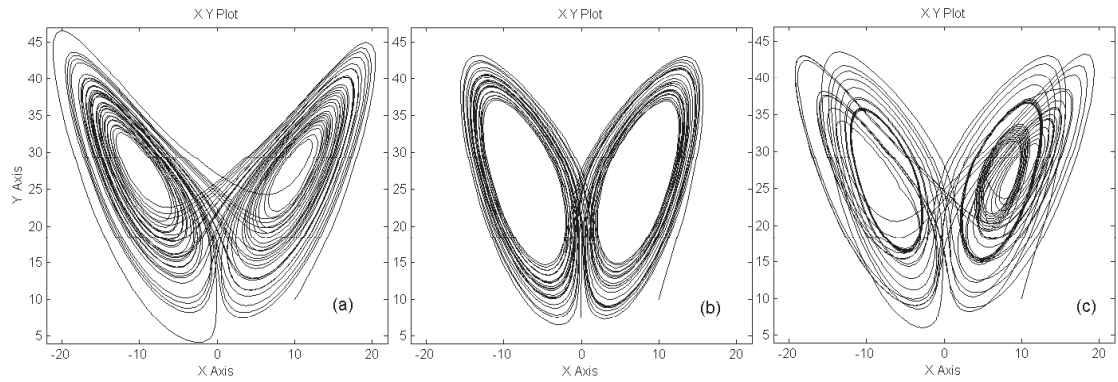}
  \caption{Lorenz attractor with different parameter values: (a)
$\sigma_{1}=7.5$; (b) $\sigma_{1}=12.5$, (c) $\sigma_{1}$ is
switched between 7.5 and 12.5 by the plaintext.}
  \label{fig:mariposas}
\end{figure}
\end{center}

The main problem with this cryptosystem
lies on the fact that the ciphertext is an analog signal, whose
waveform depends on the system parameter values and therefore on
the plaintext signal, which modulates one parameter. Consequently,
the plaintext signal may be recovered from the transmitted signal
power. Fig.~\ref{fig:mariposas} shows the Lorenz chaotic attractor
for the different values of the parameter $\sigma_1$  proposed by
the author, making apparent the strong dependence of waveforms
from the plaintext. In Fig.~\ref{fig:mariposas} (a) and (b) the
attractor corresponding to $\sigma_{1}=7.5$ and to
$\sigma_{1}=12.5$ are shown, respectively. We can observe that the
signal amplitudes are quite different. In
Fig.~\ref{fig:mariposas}(c) the attractor trajectory corresponding
to a modulation of the $\sigma_{1}$ parameter between 7.5 and 12.5
is shown. We can observe that the resulting trajectory is the
superposition of the two preceding trajectories, although both are
clearly recognizable, allowing the easy separation of each other.

To break the system we have implemented the chaotic transmitter of
the author's example with the same parameters values and initial
conditions. The simulation is identical to the one employed in the
original example, a four order Runge-Kutta integration algorithm
in MATLAB 6. A step size of $0.001$ was used.

\begin{figure}[htbp]
%\begin{center}
  \includegraphics{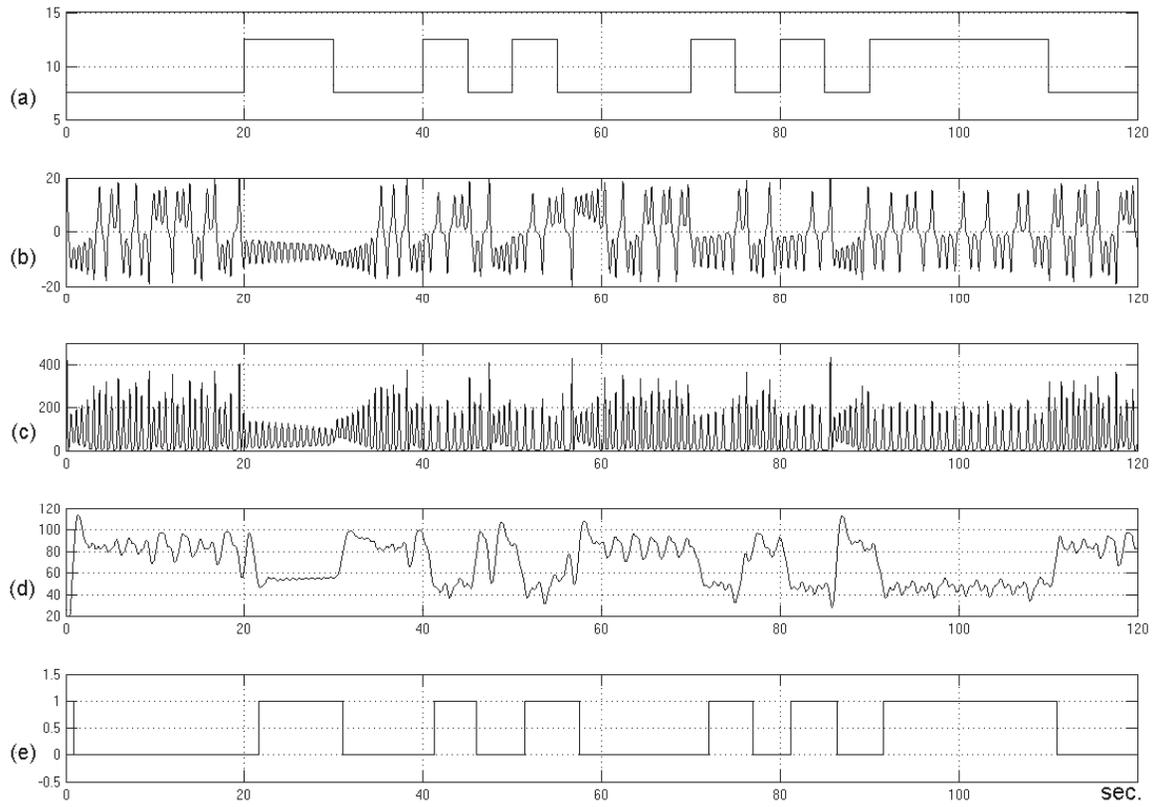}
  \caption{Power signal attack: (a) plaintext; (b) ciphertext, $x_1$; (c) squared ciphertext signal, $x_1^2$; (d) low pass
filtered
   squared ciphertext signal; (e) recovered plaintext.}
  \label{fig:power}
%\end{center}
\end{figure}

To recover the plaintext we used no chaotic receiver, instead we
computed the short time power analysis of the ciphertext. The
procedure is illustrated in Fig.~\ref{fig:power}. The first step
consists of squaring the ciphertext signal, $x_1$. Next, this
signal is low-pas filtered and, finally, binary quantized. The
low-pass filter employed is a four pole Butterworth with a
frequency cutoff of 0.5 Hz. The quantizer is an inverting
Smith-trigger with switch on point at 80 and switch off point at
50.

The result is a good estimation of the plaintext, with tiny
inaccuracies consisting of small delays in some transitions. Note
that the short initial error was also present at the beginning of
the retrieved signal obtained with the authorized receiver
described in the author's example.

It should be emphasized that our analysis is a blind detection,
made without the least knowledge of what kind of non-linear
time-varying system was used for encryption, nor its parameters
values, and neither its keys, if any.

\section{Generalized Synchronization attack}
\label{sec:GSattack}

A more precise signal retrieving of the plaintext can be performed
if we know what kind of non-linear time-varying system was used
for encryption, but still without the knowledge of its parameter
and initial condition values.

We have implemented another attack by means of an intruder
receiver based on generalized synchronization \cite{Rulkov},
fairly simpler than the authorized receiver. We use the following
receiver:

\begin{equation}\label{eq:matrix3}
\left[ {\begin{array}{*{20}c}
   {\dot {\hat {x}}_1 }  \\
   {\dot {\hat {x}}_2 }  \\
   {\dot {\hat {x}}_3 }  \\
\end{array}} \right] = \left[ {\begin{array}{*{20}c}
   { - \sigma_1} & {\sigma_2} & 0  \\
   {0} & { - 1} & 0  \\
   0 & 0 & { - b}  \\
\end{array}} \right]\left[ {\begin{array}{*{20}c}
   {\hat x{}_1}  \\
   {\hat x{}_2}  \\
   {\hat x{}_3}  \\
\end{array}} \right] + \left( {\begin{array}{*{20}c}
   0  \\
   { rx_{1}-x{}_1\hat x{}_3}  \\
   {x{}_1\hat x{}_2}  \\
\end{array}} \right)
\end{equation}

\begin{figure}[htbp]
%\begin{center}
  \includegraphics{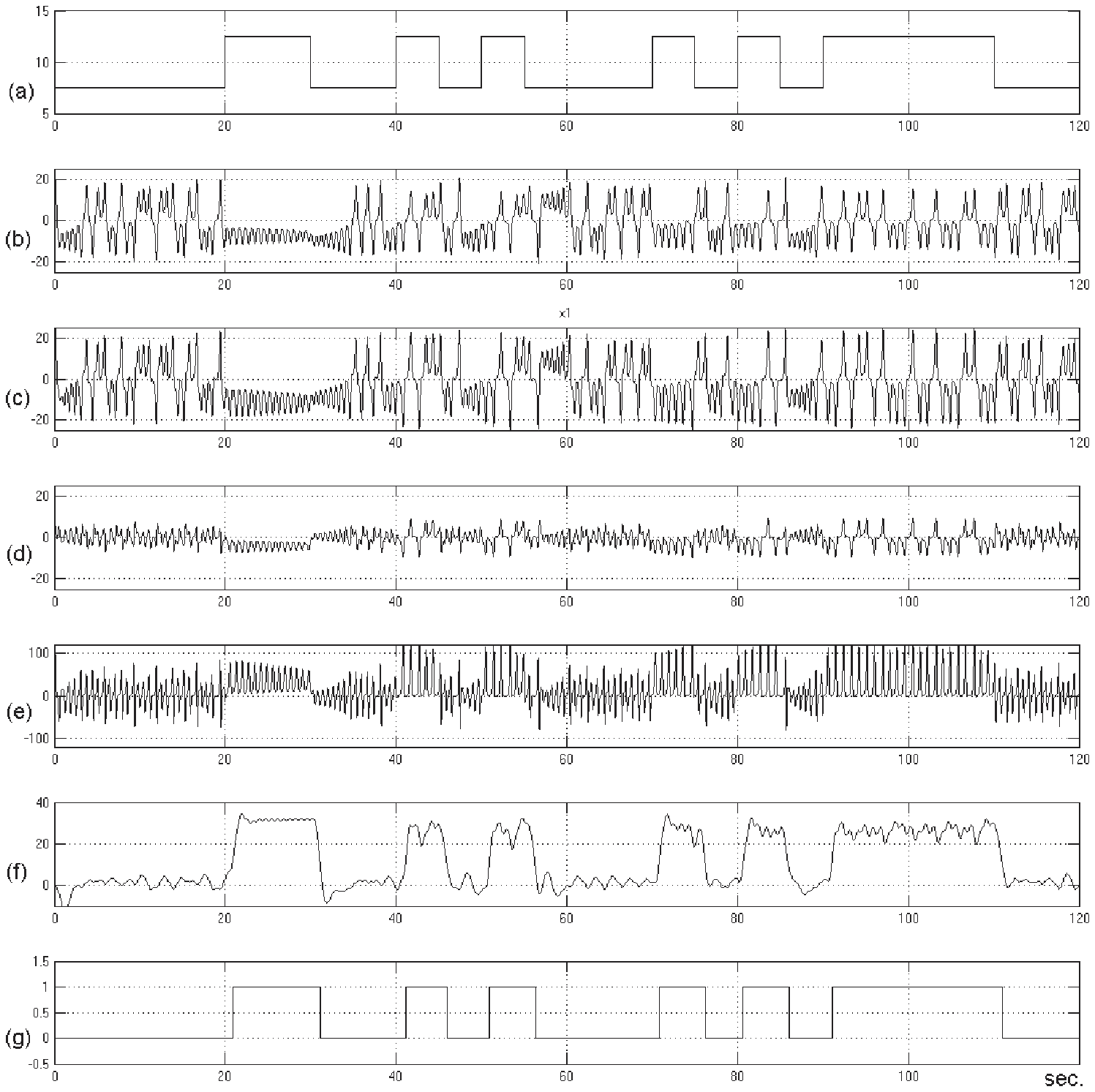}
  \caption{Generalized Synchronization attack: (a) plaintext;
  (b) ciphertext, $x_1$; (c) signal generated by the intruder's receiver, $\hat{x}_1$;
  (d) synchronization error of the intruder's receiver, $\Delta x_1=x_1-\hat{x}_1$;
  (e) ciphertext multiplied by synchronization error, $x_1\cdot\Delta x_1$;
  (f) low-pass filtering of (e); (g) recovered plaintext.}
  \label{fig:GS}
%\end{center}
\end{figure}

The plaintext recovery procedure consists of the estimation of the
short time cross correlation between the ciphertext and the
recovery error. It is illustrated in Fig.~\ref{fig:GS}. The first
step consists of calculating the synchronization error of the
receiver $\Delta x_1=x_1  - \hat x_1$. Next the synchronization
error $\Delta x_1$ is multiplied by the ciphertext $x_1$. Then
this signal is low-pass filtered. Finally, a binary quantizer is
used to regenerate the plaintext. The low-pass filter employed is
a four pole Butterworth with a frequency cutoff of 0.5 Hz. The
binary quantizer is a Smith-trigger with switch on point at 11 and
switch off point at 9.

We have found that the sensitivity to the parameter values is so
low that the original plaintext can be recovered from the
ciphertext using a receiver system with parameter values
considerably different from the ones used by the sender. The
parameter values can be obtained with a very accurate precision by
means of the trial and error procedure varying them in an effort
to approximate the filter output signal to a square wave. However,
their exact knowledge is not necessary to recover the plaintext,
as already illustrated in Fig.~\ref{fig:GS}.

The approximate range of parameter values that causes a chaotic
behavior of the Lorenz oscillator is:
\begin{align}
   \sigma_1&=\{4,14\}, \label{eq:sigma1}\\
   \sigma_2&=\{8,30\}, \label{eq:sigma2}\\
   r&=\{24,90\},\label{eq:erre}\\
   b&=\{1.5,4.5\}.\label{eq:erre}
\end{align}

Actually, we have selected for our implementation, represented in
Fig.~\ref{fig:GS}, the central value of each of the preceding
parameter ranges, that is: $(\sigma_1, \sigma_2,r,b)=(9, 19, 66,
3)$, with initial conditions
$(\hat{x}_1(0),\hat{x}_2(0),\hat{x}_3(0),\hat{\theta}(0))=(0,0,0,0)$.

\section{Conclusions}
The proposed cryptosystem is rather weak, since it can be broken
without knowing its parameter values and even without knowing the
transmitter precise structure. There is no mention about what the
key is, nor which is the key space, a fundamental aspect in every
secure communication system. The total lack of security
discourages the use of this algorithm for secure applications.

\ack{This work is supported by Ministerio de Ciencia y
Tecnolog\'{\i}a of Spain, research grant TIC2001-0586.}

\clearpage
\pagestyle{empty}

\section*{Figure captions}

\clearpage

\clearpage

\begin{thebibliography}{9}

\bibitem{Yang} T. Yang, A Survey of Chaotic Secure Communication
Systems. \emph{International Journal of Computational Cognition}
\textbf{2} (2004), 81--130.

\bibitem{Feki}Moez Feki, An adaptive chaos synchronization scheme
applied to secure communication. \emph{Chaos, Solitons and
Fractals} \textbf{18} (2003) 141--148.

\bibitem{Rulkov}N. F. Rulkov ,M. M. Sushckik, L. S. Tsimring and
H. D. I. Abarbanel, \emph{Generalized syncronization of chaos in
directionally coupled chaotic systems}\emph{Pys. Rev. E}
\textbf{51} (1995), 980--994.


\end{thebibliography}
\end{document}